\documentclass[traditabstract]{aa} 

\usepackage{graphicx}
\usepackage[varg]{txfonts}
\usepackage{lscape,longtable}
\usepackage{float}
\usepackage[latin1]{inputenc}
\usepackage{supertabular}
\usepackage[normalem]{ulem}

\usepackage{natbib}
\bibpunct{(}{)}{;}{a}{}{,} 
\newcommand{\FAC}{HE~1327$-$2326} 
\newcommand{\FeH}{$\mathrm{[Fe/H]}$} 
\newcommand{\BV}{$B-V$}
\newcommand{\JK}{$J-K$}
\newcommand{\deredBV}{$(B-V)_0$}
\newcommand{\deredJK}{$(J-K)_0$}
\newcommand{\FeHeq}[1]{$\mbox{[Fe/H]}={#1}$}       
\newcommand{\FeHsim}[1]{$\mbox{[Fe/H]}\sim{#1}$}   
\newcommand{\FeHlt}[1]{$\mbox{[Fe/H]}<{#1}$}       
\newcommand{\FeHle}[1]{$\mbox{[Fe/H]}\le{#1}$}     
\newcommand{\FeHgt}[1]{$\mbox{[Fe/H]}>{#1}$}       
\newcommand{\tefft}{$T_{\mbox{\scriptsize eff}}$}  
\begin{document}

\title{The stellar content of the Hamburg/ESO survey
	\thanks{Based on observations collected at Siding Spring Observatory.}}

	\subtitle{VI. The metallicity distribution of main-sequence turnoff stars
	in the Galactic halo}

\author{H. N. Li \inst{1,2,3}
	\and
		N. Christlieb \inst{2}
	\and
		T. Sch{\"o}rck \inst{2}
	\and
		J. E. Norris \inst{4}
	\and
		M. S. Bessell \inst{4}
	\and
		D. Yong \inst{4}
	\and
		T. C. Beers \inst{5}
	\and
		Y. S. Lee \inst{5}
	\and
		A. Frebel \inst{6}
	\and
		G. Zhao \inst{1}
        }

\institute{Key Lab of Optical Astronomy, National Astronomical Observatories,
		Chinese Academy of Sciences
              	A20 Datun Road, Chaoyang, Beijing 100012, China\\					
              	\email{lhn@nao.cas.cn;gzhao@nao.cas.cn}
	\and
             	Zentrum f{\"u}r Astronomie der Universit{\"a}t Heidelberg, Landessternwarte,
		K{\"o}nigstuhl 12, 69117 Heidelberg, Germany\\						
		\email{N.Christlieb@lsw.uni-heidelberg.de}
	\and
		Graduate University of Chinese Academy of Sciences, Beijing 100080, China		
	\and
	     	Research School of Astronomy and Astrophysics, Australian National University,
		Cotter Road, Weston, ACT 2611, Australia\\						
	     	\email{jen/bessell@mso.anu.edu.au}
	\and
		Department of Physics and Astronomy, and JINA: Joint Institute for      		
     		Nuclear Astrophysics, Michigan State University, E. Lansing, MI 48824, USA;
		\email{beers@pa.msu.edu}
	\and
		Harvard-Smithsonian Center for Astrophysics, Cambridge, MA 02138, USA;
		\email{afrebel@cfa.harvard.edu}
             }

\date{Received  / Accepted}


\abstract{We determine the metallicity distribution function (MDF) of
the Galactic halo based on metal-poor main-sequence turnoff-stars
(MSTO) which were selected from the Hamburg/ESO objective-prism survey
(HES) database.  Corresponding follow-up moderate-resolution
observations (R $\backsim$ 2000) of some 682 stars (among which 617
were accepted program stars) were carried out with the 2.3m telescope
at the Siding Spring Observatory (SSO).  Corrections for the survey
volume covered by the sample stars were quantitatively estimated and
applied to the observed MDF. The corrections are quite small, when
compared with those for a previously studied sample of metal-poor
giants.  The corrected observational MDF of the turnoff sample was
then compared with that of the giants, as well as with a number of
theoretical predictions of Galactic chemical evolution, including the
mass-loss modified Simple Model.  Although the survey-volume corrected
MDFs of the metal-poor turnoff and the halo giants notably differ in
the region of \FeHgt{-2.0}, below \FeHsim{-2.0}, (the region we
scientifically focus on most) both MDFs show a sharp drop at
\FeHsim{-3.6} and present rather similar distributions in the
low-metallicity tail.  Theoretical models can fit some parts of the
observed MDF, but none is found to simultaneously reproduce the peak
as well as the features in the metal-poor region with \FeH ~between
$-2.0$ to $-3.6$.  Among the tested models only the GAMETE model, when
normalized to the tail of the observed MDF below \FeHsim{-3.0}, and
with $Z_{cr}=10^{-3.4}Z_{\odot}$, is able to predict the sharp drop at
\FeHsim{-3.6}.

   }

\titlerunning{Metallicity Distribution of Halo Turnoff Stars}
\authorrunning{Li et al.}

\maketitle


\section{Introduction}
\label{sec:intro}
	
The Galactic halo provides important clues for understanding the evolution and
structure of the Galaxy. In the past few decades, considerable observational and
theoretical efforts have been made to investigate its chemical evolution, details
of its structure, and its kinematical characteristics. Very metal-poor stars in
the halo, those with metallicity \FeH \footnote{The common notation of
$\mathrm{[A/B]}$ = log$(N_{\mathrm A}/N_{\mathrm B}) _{\star}-$ log$(N_{\mathrm
A}/N_{\mathrm B}) _{\sun}$ is used here, where $N_{\mathrm A}$ and $N_{\mathrm
B}$ are the number densities of elements A and B, respectively.} $\lesssim
-2.0$, are regarded as fossils of the earliest generations of stars. They
preserve the chemical information created by their stellar progenitors, providing
fundamental insights regarding the properties of the very first generation of stars, the
chemical history of our Galaxy (and other large spirals like it), the modes of
star formation in the proto-Milky Way, the formation of the Galactic halo,
and physical mechanisms such as feedback processes in the early stages of galaxy
evolution. Although we are gaining a deeper understanding over time,
much remains to be explored. It is particularly revealing that,
after many decades of assuming that the Galactic halo comprises
a single stellar population, recent work \citep{Carollo2007Nature,Carollo2010ApJ}
has provided additional support to suspicions that emerged from previous efforts
that the halo is indeed divisible into two structural components, with
notably different spatial density profiles, stellar orbits, and stellar metallicities.

Recently, new theoretical models (e.g.,
\citealt{Helmi2008AARv,Prantzos2008AA, Salvadori2010MNRAS}) and
observational constraints (e.g.,
\citealt{Carollo2007Nature,Carollo2010ApJ,Bell2008ApJ,Ivezic2008ApJ,
Juric2008ApJ,Bond2009astroph,deJong2010ApJ}) have greatly enhanced our
understanding of the nature of the halo components of our Galaxy,
Those are enabling the development of plausible assembly histories
based on the degree of detectable spatial and phase-space
substructures.  The possible association of at least some presently
observed dwarf galaxies with the formation of the halo populations, as
invoked by \citet{Carollo2007Nature} to account for their dual halo
structure, has received additional support based on high-resolution
spectroscopic analysis of individual stars in ultra-faint and dwarf
spheroidal  galaxies(e.g.,
\citealt{Munoz2006ApJ,Kirby2008ApJ,Geha2009ApJ,
Frebel2010Nature,Norris2010ApJ}). Finally, the identification and
detailed analysis of the elemental abundance patterns for the most
chemically primitive stars, e.g., the ultra (\FeHlt{-4.0};
\citealt{Norris2007ApJ}) and hyper (\FeHlt{-5.0};
\citealt{Christlieb2002Nature,Frebel2005Nature,Aoki2006ApJ})
metal-poor stars allow one to trace back close to the very beginning
of star formation in the Galaxy.

The observed metallicity distribution function (MDF) of halo stars
provides strong constraints on models for the formation and chemical
evolution of the Galaxy. Any accepted model must be able to predict
the relative numbers of halo stars as a function of their metallicity
\citep{Beers2005ARAA,Helmi2008AARv}, and in the case of a dual-halo
model, as a function of location and kinematics. Early investigations
on the shape of the halo MDF were hampered by the small numbers of
very metal-poor stars known at the time
\citep{Hartwick1976ApJ,Bond1981ApJ, Ryan1991AJ,Carney1996AJ}. Other
attempts (e.g., \citealt{Bonifacio2000AJ,Schuster2004AA}), based on
samples of metal-poor stars from the HK survey of Beers and colleagues
\citep{Beers1985AJ,Beers1992AJ}, suffer from poorly constrained
selection criteria, except perhaps at the lowest metallicities. More
recent efforts have made use of statistically well-understood
selection criteria to identify large numbers of metal-poor candidates
from objective-prism surveys, such as the Hamburg/ESO survey
\citep[HES --][]{Wisotzki1996AA}, as reported in a series of papers
\citep{Barklem2005AA,HESstellarIV,HESstellarV,Placco2010AJ}.
\citet{HESstellarV}, for example, used a sample of 1638 metal-poor
giants to study the shape of the low-metallicity tail of the halo MDF,
and made detailed comparisons with MDFs of Galactic globular clusters
and satellite galaxies, as well as with theoretical models.

Main-sequence turnoff (MSTO) stars have long been used to explore
Galactic structure, including the recognition of stellar substructures
in the Galactic halo \citep{Majewski2004ApJ,An2009ApJ}, searches for
kinematic streams (e.g., \citealt{Klement2009ApJ}), and statistical
analyses of the amount of cold halo substructure in the Milky Way
(e.g., \citealt{Schlaufman2009ApJ}).  In addition, MSTO stars have
also been proven important to the field of Galactic chemical
evolution, through the analysis of high-resolution, high
signal-to-noise spectroscopic observations to derive elemental
abundances for metal-poor dwarf stars \citep{Cohen2004ApJ}, chemically
interesting metal-poor turnoff stars \citep{Aoki2008ApJ}, and
investigations of the so-called Spite Plateau \citep{spite1982AA}
through Li abundance measurements for metal-poor turnoff
stars\citep{Aoki2009ApJ,Sbordone2010astroph}.  In this paper we
construct the MDF of Galactic halo MSTO stars based on follow-up
moderate-resolution ($R \sim 2000$) spectroscopic observations of
candidate metal-poor turnoff stars from the HES.  We also compare our
results with the previously derived MDF of HES giants, and with theoretical
expectations.

This work is a continuation of the HES stellar content series
(Paper~I -- \citealt{HESstellarI}, II -- \citealt{HESstellarII},
III -- \citealt{HESstellarIII}, IV -- \citealt{HESstellarIV},
V -- \citealt{HESstellarV}). We describe the selection of the
HES turnoff sample in Section~\ref{sec:sample}, with details of
the metallicity determination and MDF construction given
in Section~\ref{sec:obs_analysis}. The observed HES MSTO MDF is
compared with theoretical predictions in Section~\ref{sec:compare_models};
the main results are summarized in Section~\ref{sec:summary}.


\section{The Sample}
\label{sec:sample}

Adopting the methods described in Paper~IV, the HES metal-poor turnoff
candidates were selected from the HES objective-prism database, using
both KP/\deredBV ~and KP/\deredJK ~selections. The only exception was
that an additional \deredBV ~range was specified so that the
candidates were restricted to 0.3 $\le$ \deredBV $\le$ 0.5.  The KP
index hereby measures the strength of the \ion{Ca}{ii}~K line and is
defined in detail in \citet{Beers1999AJ}). It was measured in all HES
prism spectra and together with a color, the prime indicator for the
selection of metal-poor candidates.

As shown in Figure 6 of Paper~IV, the employed KP cutoff becomes
comparable to its measurement uncertainty for stars within our
\deredBV ~range. Since turnoff stars are also relatively weak-lined,
we thus included only those candidates whose \ion{Ca}{ii}~K line is
not significantly detected in HES spectra, even if their KP indices
are above the formal cutoff line.  To maintain relatively consistent
exposure times during the follow-up observations, an additional cutoff
regarding the brightness of $B_{HES} \le 16.5$ was adopted.  The above
cuts yielded a preliminary sample of 3383 metal-poor turnoff
candidates from the HES database.

In order to provide candidates with a higher likelihood of being metal-poor,
the HES prism spectra of the selected 3383 candidate were visually inspected.
As defined in Paper~IV, based on the apparent strength of the \ion{Ca}{ii}~K line
relative to the continuum, the 3383 candidates were classified into
four different metal-poor classes, mpca, unid, mpcb, and mpcc.
The distributions of the 3383 candidates for these classes
are listed in the second column of Table~\ref{tab:mpclass}.

For accurate measurements of stellar metallicities (as well as the
estimates of other stellar atmospheric parameters),
moderate-resolution follow-up spectra are required. To avoid possible
systematic offsets of spectral features that could arise from
combining different telescope/detector combinations, we exclusively
adopted data observed during 15 individual runs at the Siding Spring
Observatory (SSO) 2.3m telescope with the Double Beam Spectrograph
(DBS). The runs took place between January 2006 and November 2009. The
resolving power was R $\backsim$ 2000, with a typical S/N of 20/1 per pixel
in the continuum region close to the \ion{Ca}{ii}~K line.

For a total of 682 unique stars from our metal-poor turnoff candidate
list follow-up spectra were obtained. The third column of
Table~\ref{tab:mpclass} lists the numbers of these observed candidates
for all four metal-poor classes. It is clear that a significant bias
against the class mpcc exist because it is the subjectively least
promising candidate class for finding metal-poor stars.

\begin{table}[htbp]
\caption{Numbers of different metal-poor classes among the candidate HES turnoff sample,
	the sample with follow-up observations from SSO, and the accepted sample after the visual inspection
	and rejection as described in Section~\ref{subsec:measurement}.
	The column labelled "Factor" refers to the scaling factor
	used to construct the MDF described in Section~\ref{subsec:measurement}}
\label{tab:mpclass}
\centering
\begin{tabular}{lcccc}
\hline\hline
	Class & HES prism & SSO follow-up & Accepted & Factor \\
\hline
  	mpca & 179 & 36 & 29 & 6.17 \\
	unid & 333 & 67 & 59 & 5.64 \\
	mpcb & 1666 & 560 & 513 & 3.25 \\
	mpcc & 1205 & 19 & 16 & 75.31 \\
\hline
	Total & 3383 & 682 & 617 & \\
\hline\hline
\end{tabular}
\end{table}

\section{Analysis of the Observational Sample}
\label{sec:obs_analysis}

\subsection{Measurements of line indices and [Fe/H]}
\label{subsec:measurement}

Line indices \citep{Beers1999AJ} were measured for all 682 stars in
our program sample which include the KP index, the HP2 index (which
measures the strength of the Balmer $H_{\delta}$ line), and the GP
index (which measures the strength of the of CH G-band feature).  In
the cases where we had multiple spectra for a single object, we
adopted a S/N-weighted average of the individual indices. Following
this step, a visual inspection of the follow-up spectra was carried
out to identify and reject spectra of objects that were too noisy, had
emission lines present, or were too hot (as indicated by their Balmer
line index); a few additional objects that turned out to be galaxies
or were otherwise peculiar were rejected as well. No stars with GP $>$
6\,{\AA}, which indicates strong spectral carbon features, were
detected in our sample. This is perhaps not surprising, given the
relatively high effective temperatures of our turnoff sample.  We note
that \FAC, the most metal-deficient star currently known, was one of
the candidates observed during the 15 runs. On the grounds that the
star was known to be hyper-metal-poor and in order to obtain a better
medium-resolution spectrum than previously existed, it was included in
the follow-up observations described here. However, we removed it from
our sample because including this star might introduce a bias to the
sample.  This left us with an ``accepted'' sample containing 617 stars, with
the distribution across different metal-poor classes given in Column
4 of Table~\ref{tab:mpclass}.

To obtain \FeH ~estimates for the stars in our sample, two independent
procedures were carried out.  The first obtains \FeH ~from the
measured KP and HP2 indices by using an updated code version making
use of the methods described by \citet{Beers1999AJ} (which includes
more calibration stars, and thus results in a better coverage of
stellar parameter space, especially in the lowest metallicity regime).
This was the method used in Paper V when constructing the giant-star
sample. The second method is a newly developed version of the SEGUE
Stellar Parameter Pipeline \citep[SSPP -
][]{Lee2008AJ_SSPP1,Lee2008AJ_SSPP2,Allende2008AJ_SSPP3}.
The SSPP is the software tool used to obtain estimates of atmospheric
parameters for stellar spectra obtained during the course of the Sloan
Digital Sky Survey \citep{York2000AJ,Abazajian2009ApJS} and its
extensions, SDSS-II and SDSS-III.  Recent experiments with spectral
data with similar resolving power to SDSS spectra ($R = 2000$) have
indicated that the SSPP can provide useful estimates of parameters for
non-SDSS data as well, as long as the wavelength coverage extends from
roughly 3800\,{\AA} to 5200\,{\AA}.  Slightly smaller wavelength
ranges can still be used, but the accuracy of the derived parameters
(in particular surface gravity) begins to suffer when the red limit is
less than 5000\,{\AA}, due to the loss of the Mg~I$\mathrm{b}$ and MgH
features, which provide enhanced sensitivity to estimates of $\log$ g.

The non-SEGUE Stellar Parameter Pipeline (n-SSPP) takes as inputs
user-supplied measurements or estimates of the Johnson $V$ magnitude
and \BV ~color, and/or a 2MASS \citep{Cutri2003_2MASS} $J$ magnitude
and \JK ~color, all corrected for a user-specified level of absorption
and reddening, along with a user-supplied estimate of the observed
radial velocity.  It then proceeds to determine estimates of the
primary atmospheric parameters (\tefft, $\log$ g, \FeH) and their
estimated errors, as well as estimates of distance, making use of a
subset of the procedures described in \citet{Lee2008AJ_SSPP1} (i.e.,
those that can be made to work within the wavelength region covered by
the input spectrum).  Note that it is not necessary that the input
spectra be flux calibrated, nor continuum rectified. It is also not
strictly necessary to supply input colors, since the n-SSPP makes
internal estimates that can be used as needed, but due to possible
degeneracies in the derived parameters color information is certainly
preferred.

The n-SSPP was used to obtain atmospheric parameter estimates for our
617 accepted program stars. Input Johnson colors were taken from the
estimates provided by the HES catalog (for $V$ and \BV), or
photometric measurements for a small subset of the data available from
\citet{Beers2007ApJS} or later (Beers et al., in prep.); $J$ and \JK
~were taken from the 2MASS Point Source Catalog, absorption corrected
or de-reddened according to the \citet{Schlegel1998ApJ} dust maps.

\begin{figure}[htbp]
\hspace{-0.5cm}
\includegraphics[clip=true,scale=0.55]{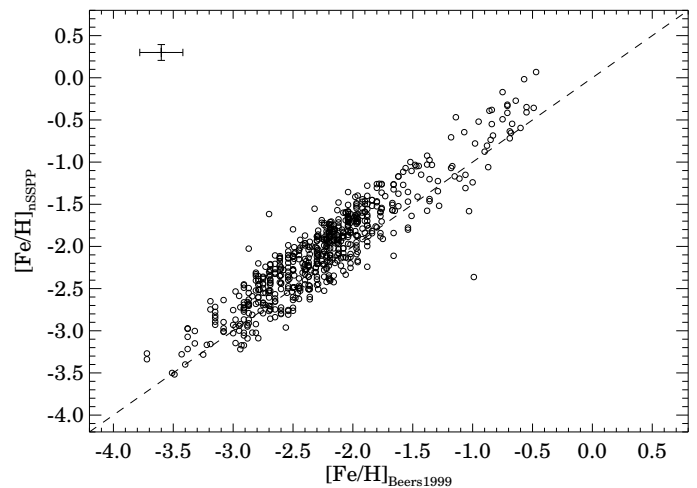}
\caption{Comparison of the metallicities (\FeH) of the HES turnoff
	sample measured with the updated version of the
	\citet{Beers1999AJ} method and the n-SSPP.  The dashed line is
	the one-to-one line; typical errors for both measurements are
	shown in the upper-left corner.  A slight offset between the
	methods is clearly present. See text for discussion.}
\label{fig:compare_FeH}
\end{figure}

The resulting estimates of \FeH ~for the two methods are compared
in Figure~\ref{fig:compare_FeH}. Although the two measurements do not
greatly differ, the typical error of the \citet{Beers1999AJ}
determination (0.18\,dex) is twice that of the n-SSPP estimate
(0.09\,dex).  The determination based on \citet{Beers1999AJ} exhibits
an offset in \FeH ~compared with the n-SSPP of $-0.19\pm0.01$,
resulting in a metallicity distribution that reaches apparently lower
\FeH.  In order to make the derived metallicity distribution as
accurate as possible we have adopted the high-resolution
measurements for the three candidates that have been observed with
high-resolution spectroscopy \citep{Cohen2004ApJ}.  As shown in
Table~\ref{tab:highres_feh}, the metallicities obtained by the n-SSPP
for these stars are closer to the values derived by the
high-resolution analysis than are those from the Beers et al.
approach. Considering the fact that the n-SSPP procedure delivers what
is likely to be a more accurate estimate of \FeH, the following
statistical discussion will be based on this method.

\begin{table}[htbp]
\caption{Comparisons of difference measurement techniques for \FeH. Columns 2-4 respectively refer to values
	derived by the updated version of the \citet{Beers1999AJ} method, the n-SSPP,
	and detailed analysis by \citet{Cohen2004ApJ} based on high-resolution spectroscopy.}
\label{tab:highres_feh}
\centering
\begin{tabular}{lccc}
\hline\hline
HES ID & $\mbox{[Fe/H]}_1$ & $\mbox{[Fe/H]}_2$ & $\mbox{[Fe/H]}_3$ \\
\hline
 HE 0007-1832 & $-$3.32 & $-$2.98 & $-$2.65 \\
 HE 0105-2202 & $-$2.87 & $-$2.86 & $-$2.55 \\
 HE 1346-0427 & $-$3.32 & $-$3.57 & $-$3.40 \\
\hline\hline
\end{tabular}
\end{table}

\begin{figure}[htbp]
\hspace{-1cm}
\includegraphics[clip=true,scale=0.55]{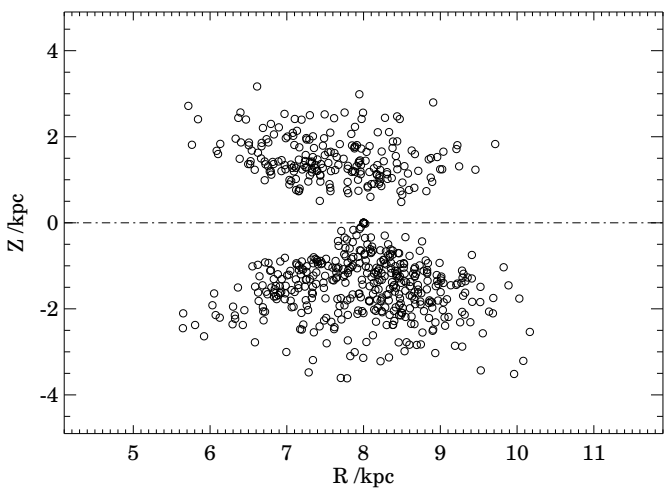}
\caption{Spatial distribution of the observed HES MSTO sample.  R indicates
	the Galactocentric distance projected to the plane; Z indicates the distance
	above or below the Galactic plane. In this diagram, the Sun is assumed to be
	located at R = 8\,kpc, Z = 0.}
\label{fig:space_dis}
\end{figure}

Distances to our sample stars were calculated by assigning the stars
into various luminosity classes. This was done based on the surface
gravity estimates derived by the n-SSPP. We considered all stars with
estimated log g $\ge 3.5$ to be likely dwarfs, those with 3.5 $<$ log
g $\le$ 3.0 to be turnoff stars, and those with log g $<$ 3.0 to be
subgiants and giants.  The distribution of the sample in the Z-R plane
is shown in Figure~\ref{fig:space_dis}. It indicates that our sample
of turnoff candidates is indeed located within $2-3$\,kpc of the Sun.
This can be contrasted with a similar figure (Figure 3) from Paper V
which shows that the giants are located at much larger distances from
the Sun.

\subsection{Observational biases and selection effects}
\label{subsec:bias}

Before comparing our observed MDF with other results or
theoretical predictions, it is necessary to address the biases and
selection effects that are introduced through the survey itself or
by our sample selection procedure.

\begin{figure*}[htbp]
\centering
\includegraphics[clip=true,scale=0.65]{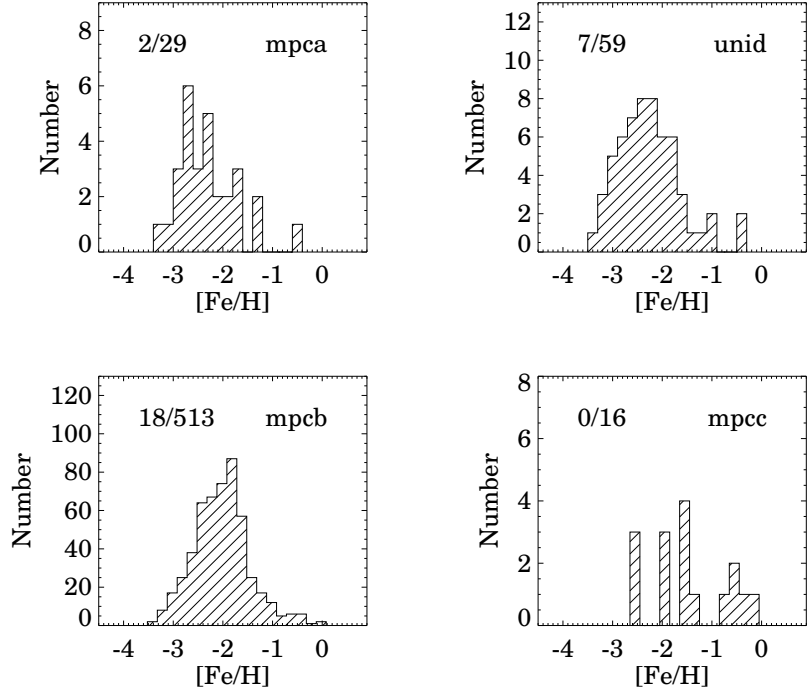}
\caption{The observed MDFs of the different metal-poor classes.
	The numbers listed in each panel correspond to the number
	of EMP stars (the numerator) compared to the total number of
	stars in the corresponding class (the denominator).}
\label{fig:mpclass}
\end{figure*}

One notable selection bias comes from the metal-poor
classification. As shown in Figure~\ref{fig:mpclass} the follow-up
observations clearly favor the best metal-poor candidates. The numbers
in the panels indicate that the relative selection efficiency of
extremely metal-poor (EMP) stars (\FeHle{-3.0}) obviously decreases
from a maximum of about 10\% for the better candidate classes, mpca
and unid, to the least likely class, mpcc. Note that the fraction of
EMP stars in class mpca is somewhat lower than that in class unid,
probably due to the fact that turnoff metal-poor stars are rather
weak-lined, making the divisions between these classes rather
difficult.  The numbers of targets in mpcc is rather small (16),
hence it is perhaps not surprising that no EMP stars were found in
this class.

\subsection{Construction of the MDF}
\label{subsec:HES_mdf}

As discussed above, the MDF derived from our follow-up observations contains a
significant bias towards the more metal-deficient candidates,
and must be taken into account to recover a reasonable
representation of the ``true'' MDF. Therefore, we adopted the scaling
factor procedure described in Paper V. For each metal-poor class, the MDF of the
observed candidates is scaled by a factor calculated from the division of the
total number in the class by the observed number (as listed in the last column
in Table~\ref{tab:mpclass}). Then the scaled MDFs of the four classes are
co-added to produce a general MDF for the entire HES candidate sample.
Similarly to Paper~V, the main difference between the directly observed and
the scaled MDF is the increasing ratio of the relatively metal-rich stars
in the mpcb and mpcc classes. The normalized fraction of the scaled MDF
is listed in the first column of Table~\ref{tab:TO_volume_MDF}.

\subsection{Selection fraction}
\label{subsec:selfrac}

\begin{figure*}[htbp]
\centering
\includegraphics[clip=true,scale=0.6]{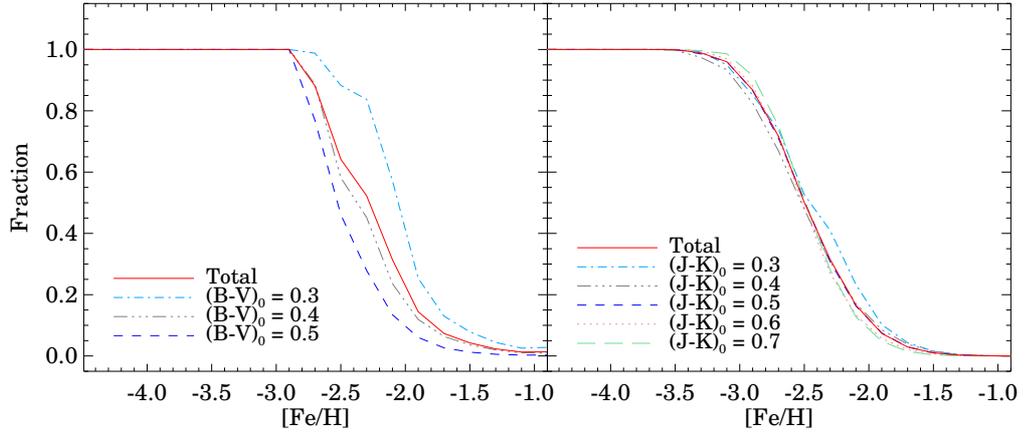}
\caption{Selection fractions of metal-poor candidates selected with the HES selection criteria
	as described in Paper~IV. The two panels correspond to selection efficiencies
	using KP and \deredBV ~(left) or \deredJK ~(right).
	Different lines refer to different red cutoffs as shown in the legend;
	the solid lines refer to the total selection fractions.}
\label{fig:selfrac}
\end{figure*}

As pointed out in Paper~IV and V, the combination of the KP index with
\deredBV~ or \deredJK~ for the purpose to select metal-poor candidates
in the HES has proven rather efficient. Following the metallicity
distribution predicted by the Simple Model, we apply our quantitative
selection criteria to a simulated sample of metal-poor stars.  The
results of the theoretical selection fractions shown in
Figure~\ref{fig:selfrac}. The selection fractions for both \deredBV
~and \deredJK~ are shown.  It is clear that the selection criteria are
able to reject the majority of stars with \FeH ~greater than
$-2.0$. For both colors, a high completeness (up to almost 100\%) is
reached for stars with \FeHle{-3.0}. For \deredBV, the redder
candidates exhibit a larger selection fraction (due to less
contamination from hot stars among the bluer candidates). The
selection fraction, however, does not differ much among the different
\deredJK~ cutoffs.  This is as expected since the blue cutoff in
\deredJK~is already fairly red so that fewer hot candidates enter the
sample.

\subsection{Survey volume correction}
\label{subsec:survey_volume}

\begin{figure*}[htbp]
\centering
\includegraphics[clip=true,scale=0.6]{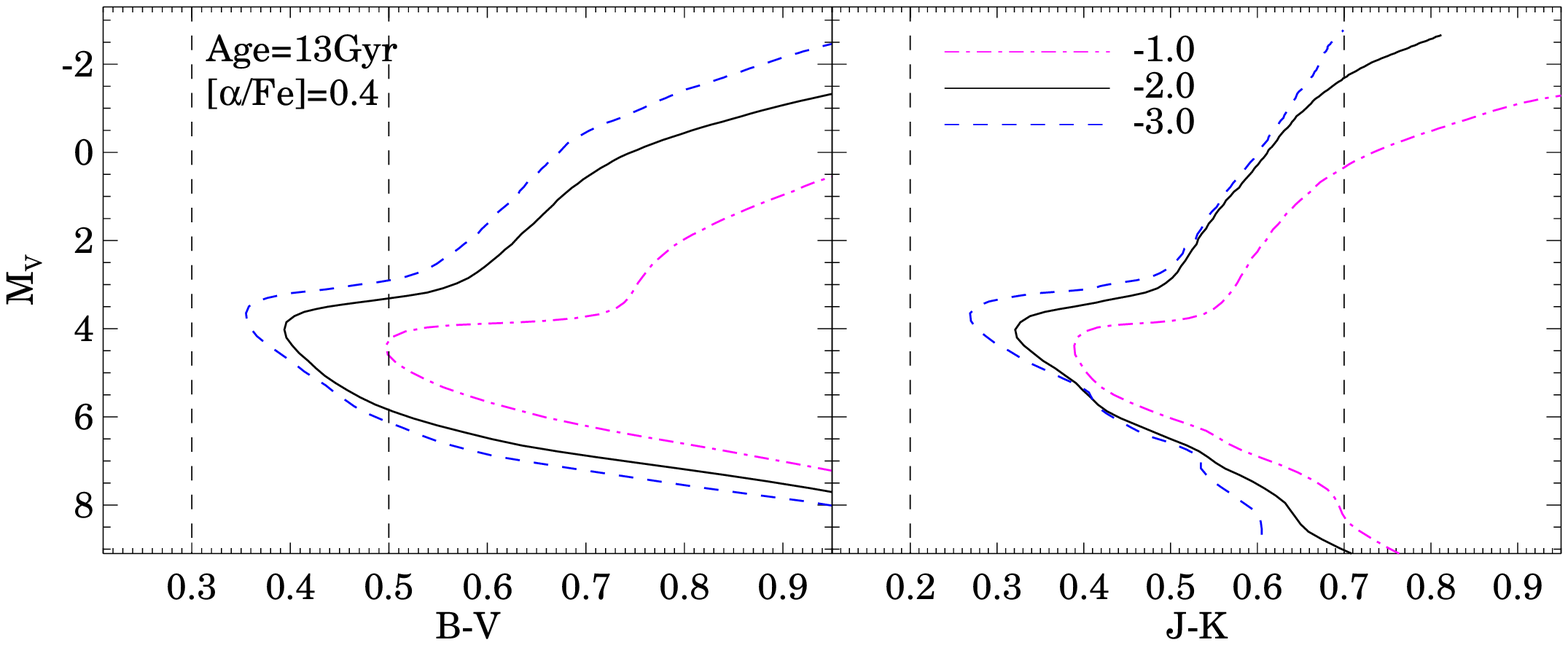}
\caption{Theoretical isochrones from \citet{Demarque2004ApJS},
	based on the Yonsei-Yale isochrones Version~2 \citep{Yi2001ApJS,
	Demarque2004ApJS}, with parameters shown in the left panel; the
	dash-dotted, solid, and dotted lines correspond to \FeHeq{-1.0}, $-2.0$, and
	$-3.0$, respectively. The color ranges of the HES turnoff sample are indicated
	by the vertical dashed lines.}
\label{fig:YYiso}
\end{figure*}

As pointed out in Paper~V, for a magnitude-limited survey the relative
survey volume explored by the observed stars differs with the stars'
metallicities, which could also be readily inferred from
Figure~\ref{fig:YYiso}. Besides, as described in
Section~\ref{sec:sample} and Table~\ref{tab:mpclass}, the HES
follow-up procedure is basically a metallicity-biased survey, which
favors candidates with lower metallicities. Thus it is interesting to
investigate to what extent this effect could impact our sample and the
resulting derived MDF. Moreover, we aim at deriving a corrected MDF
that is metallicity/volume- unbiased suitable for the comparison with
other observational results and theoretical models.

The basic idea of this correction is to derive the survey volume for
stars with different metallicities, referenced to a specific
metallicity.  Here we adopt \FeHeq{-2.0}, because it is near the peak
\FeH ~of our sample, and also close to the metallicity above which we
expect the observed MDF to deviate from the ``true'' MDF due to
metallicity selection bias. It is thus convenient for later
comparisons (the choice of a different reference \FeH~ will not
strongly affect the relative fraction of each \FeH ~bin of the
corrected MDF). Based on the definition of the survey volume, the
corrected volume referenced to \FeHeq{-2.0} in a specific \deredBV
~bin can be directly estimated from
$V=10^{0.6(M_\mathrm{V}(ref)-M_\mathrm{V})}$. As for the turnoff
sample, stars within a \deredBV ~and \FeH ~bin could be either a MSTO
star or a subgiant, which obviously explore different survey
volumes. Therefore, another step in the correction is used to estimate
the ratio of the MSTO stars to subgiants in the sample.  Using the
luminosity functions from the $Y^2$ isochrones and assuming an IMF
slope of $x=1.35$ (Salpeter index), for any specific \FeH ~and
\deredBV we can obtain the number of stars per cubic parsec per
absolute magnitude interval for both the MSTO and subgiant
branches. Hence a relative density ratio of MSTO stars versus
subgiants for the sample is obtained. Given the relative number of
MSTO stars and subgiants in each \FeH ~and \deredBV ~bin, we can then
obtain the corrected number of stars within a specific \FeH ~and
\deredBV ~bin by combining the volume and the fraction corresponding
to the MSTO and subgiant stars.

\begin{figure*}[htbp]
\centering
\includegraphics[clip=true,scale=0.6]{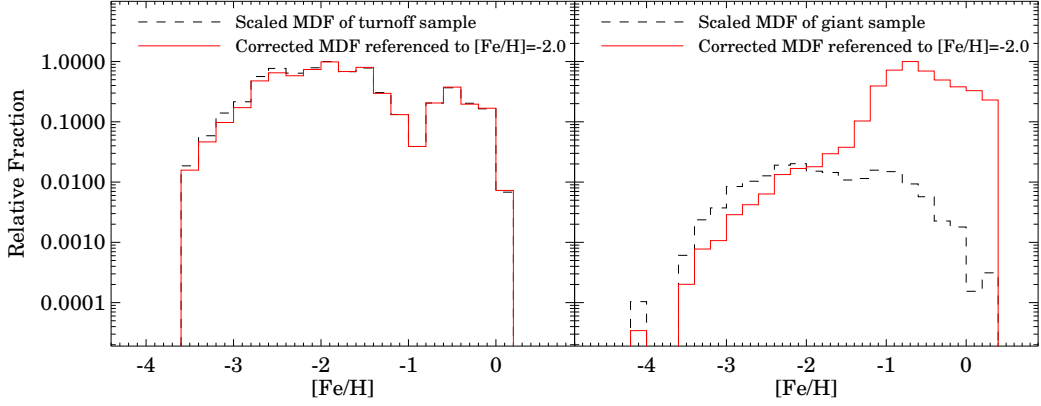}
\caption{Comparison of the observed and the survey-volume-corrected MDF,
	shown with the dashed and the solid histograms, respectively.  The left
	panel shows the HES MSTO sample, while the right panel shows the
	HES giant sample.}
\label{fig:HES_volume}
\end{figure*}

Based on this procedure, we derive the volume-corrected MDF of the
sample and compare it with the observed one. This is shown in the left panel
of Figure~\ref{fig:HES_volume}. As can be seen from inspection of this figure,
the survey-volume correction only very slightly affects the shape of the MDF. It  mildly
decreases the fraction of lower metallicity stars (referenced to \FeHeq{-2.0})
while slightly increasing the fraction at higher metallicity. This is not
unexpected because our sample of turnoff stars occupy a relatively narrow
range of \deredBV ~near the blue end of the isochrones (see
Figure~\ref{fig:YYiso}).  Thus, their relative observational volumes for
different metallicities or different branches on the isochrones (MSTO or
subgiant) do not greatly differ. This is also supported by the spatial
distribution of our sample shown in Figure~\ref{fig:space_dis}.
The correction factors of each \FeH ~bin for the MDF are listed in
the third column of Table~\ref{tab:TO_volume_MDF},
and are also applied to the corresponding \FeH ~bins of the
scaled MDF of the complete  candidate sample of 3833 stars derived in Section~\ref{subsec:HES_mdf}
(given in the last column of Table~\ref{tab:TO_volume_MDF}).

\begin{table}[htbp]
\centering
\caption{The survey-volume correction for the observed HES turnoff MDF.
	The second column refers to the fraction of the scaled MDF;
	the third column refers to the factor arising from the survey-volume effect
	as described in Section~\ref{subsec:survey_volume},
	and the last column refers to the corrected fraction for each \FeH ~bin.
\label{tab:TO_volume_MDF}}
\begin{tabular}{r|rrr}
\hline\hline
\FeH & Fraction$_0$ & Factor & Fraction \\
\hline
  $-$3.50 & 0.019 & 0.850 & 0.016 \\
  $-$3.30 & 0.059 & 0.790 & 0.047 \\
  $-$3.10 & 0.140 & 0.697 & 0.099 \\
  $-$2.90 & 0.215 & 0.799 & 0.174 \\
  $-$2.70 & 0.564 & 0.847 & 0.484 \\
  $-$2.50 & 0.770 & 0.845 & 0.659 \\
  $-$2.30 & 0.642 & 0.906 & 0.589 \\
  $-$2.10 & 0.788 & 0.937 & 0.747 \\
  $-$1.90 & 1.000 & 0.987 & 1.000 \\
  $-$1.70 & 0.679 & 1.005 & 0.691 \\
  $-$1.50 & 0.782 & 1.023 & 0.810 \\
  $-$1.30 & 0.306 & 0.970 & 0.300 \\
  $-$1.10 & 0.132 & 0.993 & 0.133 \\
  $-$0.90 & 0.039 & 0.999 & 0.039 \\
  $-$0.70 & 0.205 & 1.000 & 0.208 \\
  $-$0.50 & 0.367 & 1.027 & 0.382 \\
  $-$0.30 & 0.203 & 0.965 & 0.199 \\
  $-$0.10 & 0.164 & 1.022 & 0.170 \\
     0.10 & 0.007 & 1.069 & 0.007 \\
\hline\hline
\end{tabular}
\end{table}

To further investigate the effect of the survey-volume adjustment on the MDF,
the correction procedure was also applied to the metal-poor giant sample of Paper~V.
A similar method was adopted, except that we assumed
that the sample of Paper~V are only giants. The corrected MDF is then compared
with the observed one, as shown in the right panel of Figure~\ref{fig:HES_volume}.
The survey volume effect estimated with our method notably revises the shape of
the giants' MDF. It clearly decreases the fraction of the metal-poor component and
dramatically increases the proportion of the relatively metal-rich part.
This effect could also be expected from inspection of Figure~\ref{fig:YYiso}, because
within a certain \deredBV ~bin the survey volume explored by giants with \FeHeq{-3.0}
(when referenced to \FeHeq{-2.0}) is obviously larger than that of giants with
\FeHeq{-1.0}, resulting in a much smaller correction for
more metal-deficient giants. Thus, we conclude that although
different survey volumes for stars with different metallicities do not affect
the observed metallicity distribution of a turnoff-star dominated sample, they
will obviously change the observed MDF of a giant-dominated sample, and cannot
be ignored. In Table~\ref{tab:G_volume_factor}, we list the correction
factors for each \FeH ~bin of the giants' MDF, and applied the values to
corresponding bins of the scaled MDF of Paper~V.

\begin{table*}[htbp]
\centering
\caption{The correction factor arising from the survey-volume effect for corresponding
	\FeH ~bins of the observed giant MDF,
	as described in Section~\ref{subsec:survey_volume}\label{tab:G_volume_factor}}
\begin{tabular}{r|rrrrrrrrrrr}\hline\hline
\rule{0ex}{2.3ex} [Fe/H] & $-4.00$ & $-3.50$ & $-3.30$ & $-3.10$ & $-2.90$ & $-2.70$
                         & $-2.50$ & $-2.30$ & $-2.10$ & $-1.90$ & $-1.70$ \\
                  Factor & 0.330 & 0.331 & 0.327  & 0.287  & 0.341  & 0.407
                         &  0.499 & 0.698  & 0.834  & 1.184  & 2.048 \\\hline
\rule{0ex}{2.3ex} [Fe/H] & $-1.50$ & $-1.30$ & $-1.10$ & $-0.90$ & $-0.70$
                         & $-0.50$ & $-0.30$ & $-0.10$ & $0.10$ & $0.30$ & \\
                  Factor & 3.491  &  9.003 & 24.97 & 48.22 & 107.3
                         & 121.9 & 219.0& 211.9& 2132. & 738.4 & \\
\hline\hline
\end{tabular}
\end{table*}

\subsection{Comparison with the giants' MDF}
\label{subsec:compare_with_giant}

\begin{figure}[htbp]
\centering
\includegraphics[clip=true,scale=0.5]{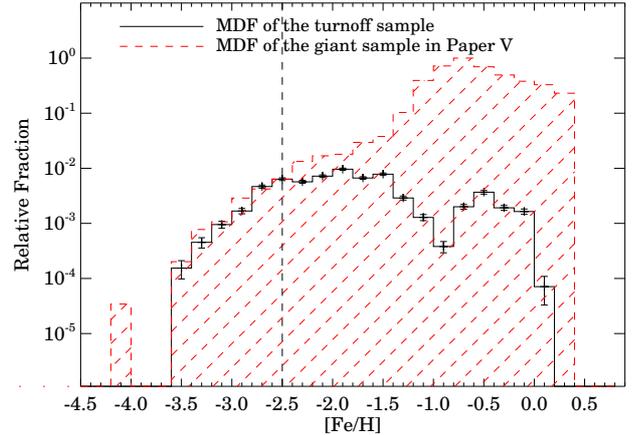}
\caption{The observed MDF of HES MSTO stars (open histogram with solid line)
	is compared with that of giants by Paper~V (filled histogram with dashed line).
	A relative normalization which scaled the maximum fraction to 1.0 is used for MDFs
	here and throughout the paper. Note that the two MDFs are scaled to
	match at \FeHeq{-2.5} (the starting point of the metal-deficient tail we defined,
    and also marked out with the vertical dashed line).}
\label{fig:HES_mdf}
\end{figure}

The MDF of the HES MSTO sample can now be compared with that of the giants from
Paper~V, as shown in Figure~\ref{fig:HES_mdf}.
The comparison between the two MDFs can be considered in two parts.

First, at the metal-poor end with \FeHlt{-2.0} (exclusive of the ultra
metal-poor component with \FeHlt{-4.0} of Paper V), both MDFs agree on the
dramatic decrease of stars below \FeHlt{-3.0} and the sharp drop at \FeHsim{-3.6}.
Besides, a $\chi^2$-test of the null hypothesis that the two samples are drawn
from the same parent distribution yields a probability of $\sim 1$. This indicates that
the two samples present are quite analogous distributions in this metallicity region.
This is not unexpected because both the turnoff and giant samples were aimed to
sample the Galactic halo population and were selected with similar criteria
in order to derive a statistically complete sample for metal-poor stars.
Thus, the two samples should follow similar statistical properties
in the metallicity region where the halo population dominates.

The two MDFs notably differ from each other in the fraction of
the relatively metal-rich component (e.g., \FeHgt{-2.0}), with the giant MDF
revealing a higher fraction. For MDFs in the region with \FeHgt{-2.0},
the $\chi^2$-test yields a probability of $\sim 0$, suggesting very different distributions.
This is not difficult to understand. As shown in the previous section,
the correction on the survey-volume has very different effects on the two MDFs.
Also, the cutoff at \deredBV $=0.5$ leads to a cutoff at comparatively lower metallicites
for the turnoff MDF. Hence, the two samples present rather distinct MDFs in this region.
However, one should keep in mind that the size of the subgroup of candidates with least possibility
of being metal-poor, i.e., mpcc, in our turnoff sample is very limited (only 16 ``accepted'' stars),
and was biased against in the whole selection and observation procedure.
Consequently, it may be incomplete for a thorough statistical comparison of MDFs in this \FeH ~region.

Therefore, as the primary motivation of this work is to discuss the properties of the halo MDF,
the completeness of both the turnoff and giant samples and the above quantitative investigation
should be reliable in the metallicity region which is of greatest interest
(\FeHlt{-2.0}, especially the metal-deficient tail between \FeHlt{-2.5} and $-3.6$).
The reader should note that the low-metallicity tail discussed in this work
is different from that discussed in Paper~V which extends to \FeHlt{-4.0}.

\section{Do Theoretical Predictions Fit the Observations?}
\label{sec:compare_models}

One of the crucial roles that the observed halo MDF  plays is to examine
and constrain theoretical models of Galactic chemical evolution. In order to
carry out such a comparison with any theoretical predictions, we first need to convert the
theoretical MDFs into a form that corresponds to what would be observed in a
survey with the same observational strategy and selection criteria as for the HES
turnoff sample.

The first modification of the theoretical MDFs is to account for the
HES selection function. To accomplish this, we inverted the
calibration of \citet{Beers1999AJ} to convert each \FeH ~into a pair
of \deredBV ~and KP or \deredJK ~and KP. Considering the fact that the
selection function varies with \deredBV ~or \deredJK ~(see Paper~IV,
V, and Section~\ref{subsec:selfrac}), these theoretical ``stars'' were
selected to follow the distribution of \deredBV ~and \deredJK ~of our
observed sample. Following this, random Gaussian errors with standard
deviations to reflect those in the measured \deredBV ~and \deredJK~
color, and the KP index were computed and added ($\sigma_{B-V}$=0.06,
$\sigma_{J-K}$=0.1, and $\sigma_{KP}$=1.0).  Finally, we applied the
same criteria for \deredBV ~or \deredJK ~versus KP to select
metal-poor ``candidates'' from these theoretical stars.  Using the
above procedure we obtain a model MDF as it would have been observed
in the HES (which we refer to as ``as observed''). We compare it with
the observed MDF of the turnoff sample in the following discussions.
Since the low-metallicity tail of the MDF is of the greatest interest
to this study, the following discussion will focus on the comparisons
in the metallicity region between \FeHeq{-2.0} and $-3.6$ (where the
observed MDF is considered statistically reliable).
  	
\subsection{Theoretical predictions based on the Simple Model}
\label{subsec:simple_model}
	
We begin our observational-theoretical comparison with the Simple Model
\citep{Searle1972ApJ,Pagel1975MNRAS} of Galactic chemical evolution.
It describes the basic form of a closed system which evolves from
initially zero-metallicity gas and remains chemically homogeneous at all times.
\citet{Hartwick1976ApJ} extended this model such that star formation ends
once the gas is either consumed or removed (essentially relaxing
the closure requirement of the system).  Here we make use of this model as
parameterized by the effective yield, $y_{\mbox{\scriptsize eff}}$,
and adopting the same value as in Paper~V, log$_{10}$ $y_{\mbox{\scriptsize eff}}=-1.7$.

\begin{figure}[htbp]
\centering
\includegraphics[clip=true,scale=0.5]{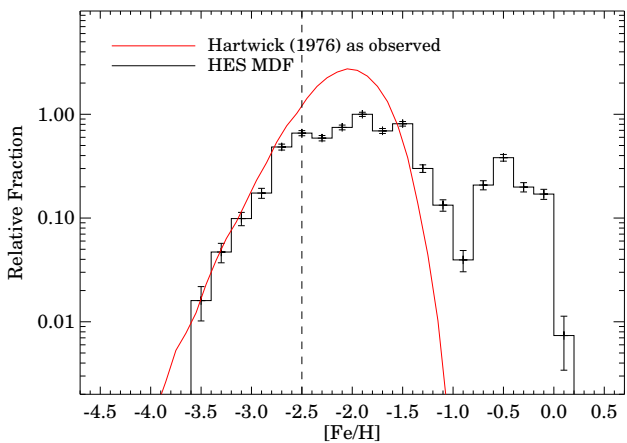}
\caption{The HES MSTO MDF (black histogram) is compared with that predicted by
	\citet{Hartwick1976ApJ}'s modified Simple Model
	(1og$_{10}$ $y_{\mbox{\scriptsize eff}} = -1.7$)
	as it would have been observed in the HES (red solid line).
	Note that for all the comparisons that follow that
	we have scaled the model MDFs in such a way that they could
    best fit the metal-poor tail of the observed MDF. The vertical dashed line
    refers to \FeHeq{-2.5} here and in the following plots too.}
\label{fig:HES_SM}
\end{figure}

The result is shown in Figure~\ref{fig:HES_SM}. As can be seen, the mass-loss
modified Simple Model is able to fit the position (\FeHsim{-2.0}) but not the height
of the peak. It does, however, well fit the general shape of MDF tail with \FeH ~from $-2.7$ through $-3.6$,
although it can only predict a smooth drop of the metal-poor tail at \FeHsim{-3.6}.
This is not entirely unexpected considering the fact that the real Galactic halo(s) could
certainly be more complicated than a simple one-zone model assuming the
Instantaneous Recycling Approximation \citep[IRA -- ][]{Tinsley1980}.

\citet{Prantzos2003AA} addressed the effect of the IRA in the determination of the MDF
of a system such as the Milky Way and suggested a physically motivated
modification to the simple outflow model, i.e., a composite model adopting a relaxed IRA,
and assuming both early infall and outflow to solve the so-called ``G dwarf problem''.
Based on this model, and the further accumulation of observational data,
\citet{Prantzos2008AA} presented a semi-analytical model in the framework
of the hierarchical merging paradigm for structure formation which assumes that
the Galactic halo is composed of the stellar debris of several sub-halos following
either the observed properties of dwarf galaxies or a structure formation calculation.

\begin{figure}[htbp]
\centering
\includegraphics[clip=true,scale=0.5]{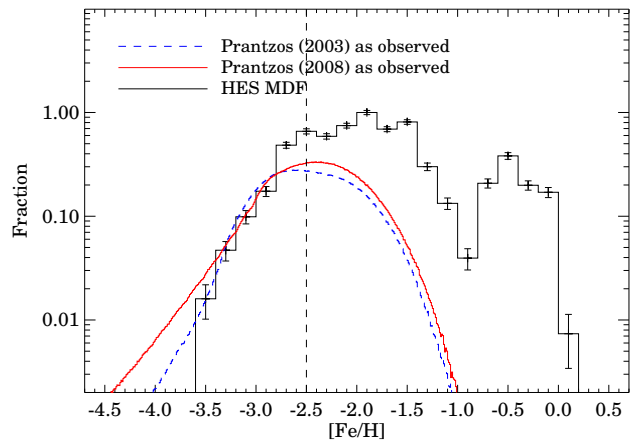}
\caption{Comparison of the HES MSTO MDF and the models of \citet{Prantzos2003AA}
	(dashed line) and \citet{Prantzos2008AA} (solid line).}
\label{fig:HES_Pt}
\end{figure}

As shown in Figure~\ref{fig:HES_Pt}, both the composite model with an
early phase of gas infall by \citet{Prantzos2003AA} and the
hierarchical merging scenario for the formation by
\citet{Prantzos2008AA} fit the shape of the observed MDF tail between
$-2.9$ and $-3.4$ rather well.  However, the location of the peak of
the MDF is not correctly predicted in either case and neither of them
reproduces the sharp drop at \FeHsim{-3.6}. Rather, they predict a
smooth decrease of numbers of EMP stars which extend to \FeHlt{-4.0}.

\subsection{Other theoretical predictions}
\label{subsec:other_model}

Besides models based on variations of the chemical evolution scheme of the
Simple Model, there are quite a number of other models based on theoretical
analyses or simulations. Here we compare our observation with two such
theoretical predictions.

\begin{figure}[htbp]
\centering
\includegraphics[clip=true,scale=0.5]{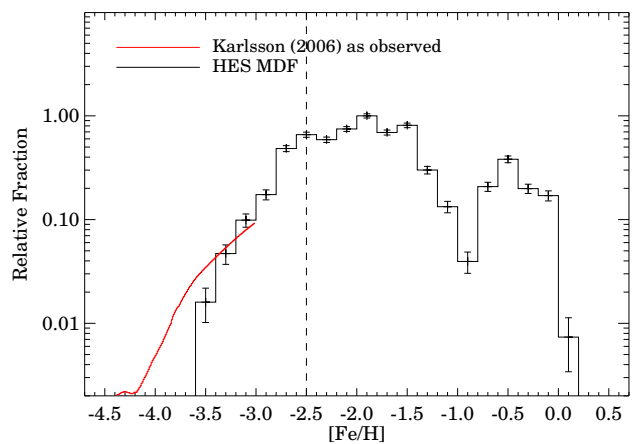}
\caption{Comparison of the HES MSTO MDF and the model of \citep{Karlsson2006ApJL}. }
\label{HES_Ks}
\end{figure}

The first considered is the model of \citet{Karlsson2006ApJL} which focuses on the
metal-poor tail with \FeHle{-3.0}, and attempts to explain the ``gap'' in the halo MDF
with \FeH ~between $-4.0$ and $-5.0$. It adoptes a scenario of negative feedback
from Population~III stars. Figure~\ref{HES_Ks} suggests that it only roughly fits
the portion of the MDF with \FeHlt{-3.0}. It also fails to predict the sharp drop at
the low-metallicity end as well.
	
\begin{figure}[htbp]
\centering
\includegraphics[clip=true,scale=0.5]{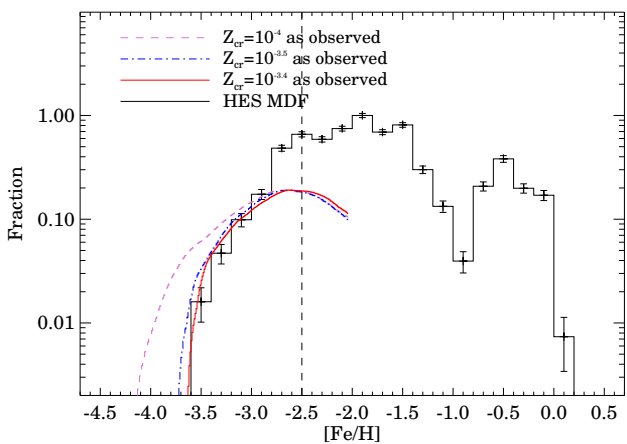}
\caption{Comparison of the HES MDF and the model of \citet{Salvadori2007MNRAS}.
	The purple dashed, blue dash-dotted, and red solid lines correspond to models with
	$Z_{cr}=10^{-4}Z_{\odot}$, $10^{-3.5}Z_{\odot}$, and $10^{-3.4}Z_{\odot}$, respectively.}
\label{HES_Zcr}
\end{figure}

Another model that has been tested is GAlaxy MErger Tree and Evolution
\citep[GAMETE -- ][]{Salvadori2007MNRAS}. It is a Monte Carlo code to reconstruct
the merger tree of the Milky Way and to follow the evolution of gas and stars along
the tree. This model defines an input parameter, the critical metallicity
$Z_{cr}$, which governs the transition from Pop~III to Pop~II star formation. We
compare our observed MDF with the simulated results corresponding to different
values of $Z_{cr}$, as shown in Figure~\ref{HES_Zcr}. Although according to the
observational data available at that time, $Z_{cr}=10^{-4}Z_{\odot}$ was
regarded as the fiducial model, it obviously cannot fit our observations here.
All the predictions fail to fit the location of the peak of the observed MDF.
Similarly to the conclusions in Paper~V, the model with $Z_{cr}=10^{-3.4}Z_{\odot}$
appears to partially fit our observed MDF, being able to reproduce the tail
with \FeHlt{-3.0} and best predict the sharp drop at \FeHsim{-3.6}.
	
\section{Summary and Discussions}
\label{sec:summary}

Based on the (for now) largest metal-poor turnoff-star sample from the HES database
and moderate-resolution follow-up observations, we have statistically
investigated the MDF of local MSTO stars in the Galactic halo.
	
\begin{enumerate}

\item With reference to \FeHeq{-2.0}, the effects of relative survey volumes
have been quantitatively estimated based on theoretical isochrones and applied to
the observed MDFs of both the HES turnoff and giants samples. It is shown that the
survey-volume effect does not substantially alter the turnoff MDF while it dramatically
changes the MDF of the giant sample from Paper~V.

\item The survey-volume corrected and metal-poor-class scaled MDFs of
the turnoff sample has been compared with that of the halo
giants. Though the two MDFs notably differ in the region with
\FeHgt{-2.0} (where our sample starts to be incomplete), for the
metal-deficient region (e.g., \FeHlt{-2.0}) the $\chi^2$-test suggests
that the two MDFs are quite similar. Furthermore, both MDFs agree regarding
the sharp drop at \FeHsim{-3.6}.  Hence, for an MDF dominated by the
halo population, the two MDFs agree well.

\item Theoretical models of Galactic chemical evolution have been
discussed. They can only fit portions of the observed MSTO MDF while
none of them fully reproduces the features of the observations. In
particular, they fail to simultaneously fit the peak and the
metal-deficient tail between \FeHsim{-2.5} to $-3.6$.  Although the
$Z_{cr}=10^{-3.4}Z_{\odot}$ case of the \citet{Salvadori2007MNRAS}
model can only partially fit the observed MDF it is able to best
predict the sharp drop at \FeHsim{-3.6}.

\item Generally, both selection criteria using KP plus \deredBV ~and \deredJK ~serve as
efficient selectors of metal-poor stars.  They are capable to reach a selection
fraction up to 100\% for the EMP candidates of our sample.

\end{enumerate}

Considering the fact that our sample mainly consists of unevolved main-sequence
(and subgiant) stars with low metallicities, it could also provide additional
useful information on Galactic chemical evolution. For example, a kinematic
analysis of this sample could be used to re-visit the role of accretion of the
interstellar medium during the long lifetimes of metal-poor stars, as
approximately calculated in a number of early works (e.g.,
\citealt{Talbot1977ApJS,Yoshii1981AA,Iben1983MmSAI}) and also discussed by more
recent studies (e.g, \citealt{Christlieb2004ApJ,Norris2007ApJ,Frebel2009MNRAS}).

It should also be pointed out that all of our comparisons of the MDFs
have been performed under the assumption that we are modeling a {\it
single} halo population, which current evidence suggests is an
over-simplification. It seems likely that the observed MDFs for both
the HES MSTO stars and the HES giants comprise overlapping
contributions from the outer-halo population at the lowest
metallicities and the inner halo at intermediate low metallicities,
with respective tails of as-yet unknown relative strengths and
convolved with the HES metallicity selection bias that becomes more
severe above \FeHsim{-2.5} to $-2.0$. This possibility was already
mentioned in Paper V where it was noted that there appeared to be
relatively larger fractions of EMP stars at heights above the plane
$|$Z$|$ $>$ 15 kpc than in the intermediate range 5 $<$ $|$Z$|$ $<$ 15
kpc. This is in line with the expectations of the dual halo
interpretation of \citet{Carollo2007Nature,Carollo2010ApJ}.  Progress
on this issue will come from consideration of the dual halo modeling
approach, ideally in combination with a full kinematic analysis of
these samples that forms the basis of a paper in preparation.

However, the HES metal-poor turnoff sample discussed in this paper
contains no objects with \FeHlt{-3.6} which obviously do exist. Thus,
we are not able to discuss the performance of theoretical MDFs in the
most metal-deficient regime. Larger statistically complete
samples are required for a thorough comparison with theoretical
predictions.  Fortunately, such samples will be obtained from much
larger and deeper surveys in the near future, such as from SEGUE-2 and
the Apache POint Galactic Evolution Experiment (APOGEE), the Large Sky
Area Multi-Object Fiber Spectroscopic Telescope (LAMOST,
\citealt{Zhao2006ChJAA}), and the Southern Sky Survey
\citep{Keller2007PASA}.

\begin{acknowledgements}

We express our sincere gratitude to the anonymous referee for the constructive comments.
H.N.L. would like to thank N. Prantzos, T. Karlsson, and S. Salvadori for
providing electronic versions of their theoretical MDF models and helpful comments.
H.N.L. and N.C. acknowledge support from the Global Networks program of
the University of Heidelberg and from Deutsche Forschungsgemeinschaft under grant CH~214/5--1.
Studies at RSAA, ANU, of the most metal-poor stellar populations are
supported by Australian Research Council grants DP0663562 and
DP0984924, which J.E.N., M.S.B., and D.Y. are pleased to acknowledge.
T.C.B. and Y.S.L. acknowledge partial funding
of this work from grants PHY 02-16783 and PHY 08-22648: Physics Frontier
Center/Joint Institute for Nuclear Astrophysics (JINA), awarded by the U.S.
National Science Foundation.
This research is partly supported by the National Natural Science Foundation
of China under grant No.10821061 and National Basic Research Program of China
(973 Program) under grant No.2007CB815103, which H.N.L. and G.Z. would like to acknowledge.
This publication makes use of data products from the Two Micron All
Sky Survey, which is a joint project of the University of Massachusetts
and the Infrared Processing and Analysis Center/California Institute of
Technology, funded by the National Aeronautics and Space Administration
and the National Science Foundation.

\end{acknowledgements}

\bibliographystyle{aa} 
\bibliography{hes_to_mdf} 

\begin{thebibliography}{69}
\expandafter\ifx\csname natexlab\endcsname\relax\def\natexlab#1{#1}\fi

\bibitem[{{Abazajian} {et~al.}(2009){Abazajian}, {Adelman-McCarthy},
  {Ag{\"u}eros}, {Allam}, {Allende Prieto}, {An}, {Anderson}, {Anderson},
  {Annis}, {Bahcall}, {Bailer-Jones}, {Barentine}, {Bassett}, {Becker},
  {Beers}, {Bell}, {Belokurov}, {Berlind}, {Berman}, {Bernardi}, {Bickerton},
  {Bizyaev}, {Blakeslee}, {Blanton}, {Bochanski}, {Boroski}, {Brewington},
  {Brinchmann}, {Brinkmann}, {Brunner}, {Budav{\'a}ri}, {Carey}, {Carliles},
  {Carr}, {Castander}, {Cinabro}, {Connolly}, {Csabai}, {Cunha}, {Czarapata},
  {Davenport}, {de Haas}, {Dilday}, {Doi}, {Eisenstein}, {Evans}, {Evans},
  {Fan}, {Friedman}, {Frieman}, {Fukugita}, {G{\"a}nsicke}, {Gates},
  {Gillespie}, {Gilmore}, {Gonzalez}, {Gonzalez}, {Grebel}, {Gunn},
  {Gy{\"o}ry}, {Hall}, {Harding}, {Harris}, {Harvanek}, {Hawley}, {Hayes},
  {Heckman}, {Hendry}, {Hennessy}, {Hindsley}, {Hoblitt}, {Hogan}, {Hogg},
  {Holtzman}, {Hyde}, {Ichikawa}, {Ichikawa}, {Im}, {Ivezi{\'c}}, {Jester},
  {Jiang}, {Johnson}, {Jorgensen}, {Juri{\'c}}, {Kent}, {Kessler}, {Kleinman},
  {Knapp}, {Konishi}, {Kron}, {Krzesinski}, {Kuropatkin}, {Lampeitl},
  {Lebedeva}, {Lee}, {Lee}, {Leger}, {L{\'e}pine}, {Li}, {Lima}, {Lin}, {Long},
  {Loomis}, {Loveday}, {Lupton}, {Magnier}, {Malanushenko}, {Malanushenko},
  {Mandelbaum}, {Margon}, {Marriner}, {Mart{\'{\i}}nez-Delgado}, {Matsubara},
  {McGehee}, {McKay}, {Meiksin}, {Morrison}, {Mullally}, {Munn}, {Murphy},
  {Nash}, {Nebot}, {Neilsen}, {Newberg}, {Newman}, {Nichol}, {Nicinski},
  {Nieto-Santisteban}, {Nitta}, {Okamura}, {Oravetz}, {Ostriker}, {Owen},
  {Padmanabhan}, {Pan}, {Park}, {Pauls}, {Peoples}, {Percival}, {Pier}, {Pope},
  {Pourbaix}, {Price}, {Purger}, {Quinn}, {Raddick}, {Fiorentin}, {Richards},
  {Richmond}, {Riess}, {Rix}, {Rockosi}, {Sako}, {Schlegel}, {Schneider},
  {Scholz}, {Schreiber}, {Schwope}, {Seljak}, {Sesar}, {Sheldon}, {Shimasaku},
  {Sibley}, {Simmons}, {Sivarani}, {Smith}, {Smith}, {Smol{\v c}i{\'c}},
  {Snedden}, {Stebbins}, {Steinmetz}, {Stoughton}, {Strauss}, {Subba Rao},
  {Suto}, {Szalay}, {Szapudi}, {Szkody}, {Tanaka}, {Tegmark}, {Teodoro},
  {Thakar}, {Tremonti}, {Tucker}, {Uomoto}, {Vanden Berk}, {Vandenberg},
  {Vidrih}, {Vogeley}, {Voges}, {Vogt}, {Wadadekar}, {Watters}, {Weinberg},
  {West}, {White}, {Wilhite}, {Wonders}, {Yanny}, {Yocum}, {York}, {Zehavi},
  {Zibetti}, \& {Zucker}}]{Abazajian2009ApJS}
{Abazajian}, K.~N., {Adelman-McCarthy}, J.~K., {Ag{\"u}eros}, M.~A., {et~al.}
  2009, \apjs, 182, 543

\bibitem[{{Allende Prieto} {et~al.}(2008){Allende Prieto}, {Sivarani}, {Beers},
  {Lee}, {Koesterke}, {Shetrone}, {Sneden}, {Lambert}, {Wilhelm}, {Rockosi},
  {Lai}, {Yanny}, {Ivans}, {Johnson}, {Aoki}, {Bailer-Jones}, \& {Re
  Fiorentin}}]{Allende2008AJ_SSPP3}
{Allende Prieto}, C., {Sivarani}, T., {Beers}, T.~C., {et~al.} 2008, \aj, 136,
  2070

\bibitem[{{An} {et~al.}(2009){An}, {Johnson}, {Beers}, {Pinsonneault},
  {Terndrup}, {Delahaye}, {Lee}, {Masseron}, \& {Yanny}}]{An2009ApJ}
{An}, D., {Johnson}, J.~A., {Beers}, T.~C., {et~al.} 2009, \apjl, 707, L64

\bibitem[{{Aoki} {et~al.}(2009){Aoki}, {Barklem}, {Beers}, {Christlieb},
  {Inoue}, {Garc{\'{\i}}a P{\'e}rez}, {Norris}, \& {Carollo}}]{Aoki2009ApJ}
{Aoki}, W., {Barklem}, P.~S., {Beers}, T.~C., {et~al.} 2009, \apj, 698, 1803

\bibitem[{{Aoki} {et~al.}(2008){Aoki}, {Beers}, {Sivarani}, {Marsteller},
  {Lee}, {Honda}, {Norris}, {Ryan}, \& {Carollo}}]{Aoki2008ApJ}
{Aoki}, W., {Beers}, T.~C., {Sivarani}, T., {et~al.} 2008, \apj, 678, 1351

\bibitem[{{Aoki} {et~al.}(2006){Aoki}, {Frebel}, {Christlieb}, {Norris},
  {Beers}, {Minezaki}, {Barklem}, {Honda}, {Takada-Hidai}, {Asplund}, {Ryan},
  {Tsangarides}, {Eriksson}, {Steinhauer}, {Deliyannis}, {Nomoto}, {Fujimoto},
  {Ando}, {Yoshii}, \& {Kajino}}]{Aoki2006ApJ}
{Aoki}, W., {Frebel}, A., {Christlieb}, N., {et~al.} 2006, \apj, 639, 897

\bibitem[{{Barklem} {et~al.}(2005){Barklem}, {Christlieb}, {Beers}, {Hill},
  {Bessell}, {Holmberg}, {Marsteller}, {Rossi}, {Zickgraf}, \&
  {Reimers}}]{Barklem2005AA}
{Barklem}, P.~S., {Christlieb}, N., {Beers}, T.~C., {et~al.} 2005, \aap, 439,
  129

\bibitem[{{Beers} \& {Christlieb}(2005)}]{Beers2005ARAA}
{Beers}, T.~C. \& {Christlieb}, N. 2005, \araa, 43, 531

\bibitem[{{Beers} {et~al.}(2007){Beers}, {Flynn}, {Rossi}, {Sommer-Larsen},
  {Wilhelm}, {Marsteller}, {Lee}, {De Lee}, {Krugler}, {Deliyannis}, {Simmons},
  {Mills}, {Zickgraf}, {Holmberg}, {{\"O}nehag}, {Eriksson}, {Terndrup},
  {Salim}, {Andersen}, {Nordstr{\"o}m}, {Christlieb}, {Frebel}, \&
  {Rhee}}]{Beers2007ApJS}
{Beers}, T.~C., {Flynn}, C., {Rossi}, S., {et~al.} 2007, \apjs, 168, 128

\bibitem[{{Beers} {et~al.}(1985){Beers}, {Preston}, \&
  {Shectman}}]{Beers1985AJ}
{Beers}, T.~C., {Preston}, G.~W., \& {Shectman}, S.~A. 1985, \aj, 90, 2089

\bibitem[{{Beers} {et~al.}(1992){Beers}, {Preston}, \&
  {Shectman}}]{Beers1992AJ}
{Beers}, T.~C., {Preston}, G.~W., \& {Shectman}, S.~A. 1992, \aj, 103, 1987

\bibitem[{{Beers} {et~al.}(1999){Beers}, {Rossi}, {Norris}, {Ryan}, \&
  {Shefler}}]{Beers1999AJ}
{Beers}, T.~C., {Rossi}, S., {Norris}, J.~E., {Ryan}, S.~G., \& {Shefler}, T.
  1999, \aj, 117, 981

\bibitem[{{Bell} {et~al.}(2008){Bell}, {Zucker}, {Belokurov}, {Sharma},
  {Johnston}, {Bullock}, {Hogg}, {Jahnke}, {de Jong}, {Beers}, {Evans},
  {Grebel}, {Ivezi{\'c}}, {Koposov}, {Rix}, {Schneider}, {Steinmetz}, \&
  {Zolotov}}]{Bell2008ApJ}
{Bell}, E.~F., {Zucker}, D.~B., {Belokurov}, V., {et~al.} 2008, \apj, 680, 295

\bibitem[{{Bond}(1981)}]{Bond1981ApJ}
{Bond}, H.~E. 1981, \apj, 248, 606

\bibitem[{{Bond} {et~al.}(2009){Bond}, {Ivezic}, {Sesar}, {Juric}, \&
  {Munn}}]{Bond2009astroph}
{Bond}, N.~A., {Ivezic}, Z., {Sesar}, B., {Juric}, M., \& {Munn}, J. 2009,
  ArXiv e-prints

\bibitem[{{Bonifacio} {et~al.}(2000){Bonifacio}, {Monai}, \&
  {Beers}}]{Bonifacio2000AJ}
{Bonifacio}, P., {Monai}, S., \& {Beers}, T.~C. 2000, \aj, 120, 2065

\bibitem[{{Carney} {et~al.}(1996){Carney}, {Laird}, {Latham}, \&
  {Aguilar}}]{Carney1996AJ}
{Carney}, B.~W., {Laird}, J.~B., {Latham}, D.~W., \& {Aguilar}, L.~A. 1996,
  \aj, 112, 668

\bibitem[{{Carollo} {et~al.}(2010){Carollo}, {Beers}, {Chiba}, {Norris},
  {Freeman}, {Lee}, {Ivezi{\'c}}, {Rockosi}, \& {Yanny}}]{Carollo2010ApJ}
{Carollo}, D., {Beers}, T.~C., {Chiba}, M., {et~al.} 2010, \apj, 712, 692

\bibitem[{{Carollo} {et~al.}(2007){Carollo}, {Beers}, {Lee}, {Chiba}, {Norris},
  {Wilhelm}, {Sivarani}, {Marsteller}, {Munn}, {Bailer-Jones}, {Fiorentin}, \&
  {York}}]{Carollo2007Nature}
{Carollo}, D., {Beers}, T.~C., {Lee}, Y.~S., {et~al.} 2007, \nat, 450, 1020

\bibitem[{{Christlieb} {et~al.}(2005){Christlieb}, {Beers}, {Thom}, {Wilhelm},
  {Rossi}, {Flynn}, {Wisotzki}, \& {Reimers}}]{HESstellarIII}
{Christlieb}, N., {Beers}, T.~C., {Thom}, C., {et~al.} 2005, \aap, 431, 143

\bibitem[{{Christlieb} {et~al.}(2002){Christlieb}, {Bessell}, {Beers},
  {Gustafsson}, {Korn}, {Barklem}, {Karlsson}, {Mizuno-Wiedner}, \&
  {Rossi}}]{Christlieb2002Nature}
{Christlieb}, N., {Bessell}, M.~S., {Beers}, T.~C., {et~al.} 2002, \nat, 419,
  904

\bibitem[{{Christlieb} {et~al.}(2001{\natexlab{a}}){Christlieb}, {Green},
  {Wisotzki}, \& {Reimers}}]{HESstellarII}
{Christlieb}, N., {Green}, P.~J., {Wisotzki}, L., \& {Reimers}, D.
  2001{\natexlab{a}}, \aap, 375, 366

\bibitem[{{Christlieb} {et~al.}(2004){Christlieb}, {Gustafsson}, {Korn},
  {Barklem}, {Beers}, {Bessell}, {Karlsson}, \&
  {Mizuno-Wiedner}}]{Christlieb2004ApJ}
{Christlieb}, N., {Gustafsson}, B., {Korn}, A.~J., {et~al.} 2004, \apj, 603,
  708

\bibitem[{{Christlieb} {et~al.}(2008){Christlieb}, {Sch{\"o}rck}, {Frebel},
  {Beers}, {Wisotzki}, \& {Reimers}}]{HESstellarIV}
{Christlieb}, N., {Sch{\"o}rck}, T., {Frebel}, A., {et~al.} 2008, \aap, 484,
  721

\bibitem[{{Christlieb} {et~al.}(2001{\natexlab{b}}){Christlieb}, {Wisotzki},
  {Reimers}, {Homeier}, {Koester}, \& {Heber}}]{HESstellarI}
{Christlieb}, N., {Wisotzki}, L., {Reimers}, D., {et~al.} 2001{\natexlab{b}},
  \aap, 366, 898

\bibitem[{{Cohen} {et~al.}(2004){Cohen}, {Christlieb}, {McWilliam}, {Shectman},
  {Thompson}, {Wasserburg}, {Ivans}, {Dehn}, {Karlsson}, \&
  {Melendez}}]{Cohen2004ApJ}
{Cohen}, J.~G., {Christlieb}, N., {McWilliam}, A., {et~al.} 2004, \apj, 612,
  1107

\bibitem[{{Cutri} {et~al.}(2003){Cutri}, {Skrutskie}, {van Dyk}, {Beichman},
  {Carpenter}, {Chester}, {Cambresy}, {Evans}, {Fowler}, {Gizis}, {Howard},
  {Huchra}, {Jarrett}, {Kopan}, {Kirkpatrick}, {Light}, {Marsh}, {McCallon},
  {Schneider}, {Stiening}, {Sykes}, {Weinberg}, {Wheaton}, {Wheelock}, \&
  {Zacarias}}]{Cutri2003_2MASS}
{Cutri}, R.~M., {Skrutskie}, M.~F., {van Dyk}, S., {et~al.} 2003, {2MASS All
  Sky Catalog of point sources.}, ed. {Cutri, R.~M., Skrutskie, M.~F., van Dyk,
  S., Beichman, C.~A., Carpenter, J.~M., Chester, T., Cambresy, L., Evans, T.,
  Fowler, J., Gizis, J., Howard, E., Huchra, J., Jarrett, T., Kopan, E.~L.,
  Kirkpatrick, J.~D., Light, R.~M., Marsh, K.~A., McCallon, H., Schneider, S.,
  Stiening, R., Sykes, M., Weinberg, M., Wheaton, W.~A., Wheelock, S., \&
  Zacarias, N.}

\bibitem[{{de Jong} {et~al.}(2010){de Jong}, {Yanny}, {Rix}, {Dolphin},
  {Martin}, \& {Beers}}]{deJong2010ApJ}
{de Jong}, J.~T.~A., {Yanny}, B., {Rix}, H., {et~al.} 2010, \apj, 714, 663

\bibitem[{{Demarque} {et~al.}(2004){Demarque}, {Woo}, {Kim}, \&
  {Yi}}]{Demarque2004ApJS}
{Demarque}, P., {Woo}, J., {Kim}, Y., \& {Yi}, S.~K. 2004, \apjs, 155, 667

\bibitem[{{Frebel} {et~al.}(2005){Frebel}, {Aoki}, {Christlieb}, {Ando},
  {Asplund}, {Barklem}, {Beers}, {Eriksson}, {Fechner}, {Fujimoto}, {Honda},
  {Kajino}, {Minezaki}, {Nomoto}, {Norris}, {Ryan}, {Takada-Hidai},
  {Tsangarides}, \& {Yoshii}}]{Frebel2005Nature}
{Frebel}, A., {Aoki}, W., {Christlieb}, N., {et~al.} 2005, \nat, 434, 871

\bibitem[{{Frebel} {et~al.}(2009){Frebel}, {Johnson}, \&
  {Bromm}}]{Frebel2009MNRAS}
{Frebel}, A., {Johnson}, J.~L., \& {Bromm}, V. 2009, \mnras, 392, L50

\bibitem[{{Frebel} {et~al.}(2010){Frebel}, {Kirby}, \&
  {Simon}}]{Frebel2010Nature}
{Frebel}, A., {Kirby}, E.~N., \& {Simon}, J.~D. 2010, \nat, 464, 72

\bibitem[{{Geha} {et~al.}(2009){Geha}, {Willman}, {Simon}, {Strigari}, {Kirby},
  {Law}, \& {Strader}}]{Geha2009ApJ}
{Geha}, M., {Willman}, B., {Simon}, J.~D., {et~al.} 2009, \apj, 692, 1464

\bibitem[{{Hartwick}(1976)}]{Hartwick1976ApJ}
{Hartwick}, F.~D.~A. 1976, \apj, 209, 418

\bibitem[{{Helmi}(2008)}]{Helmi2008AARv}
{Helmi}, A. 2008, \aapr, 15, 145

\bibitem[{{Iben}(1983)}]{Iben1983MmSAI}
{Iben}, Jr., I. 1983, Memorie della Societa Astronomica Italiana, 54, 321

\bibitem[{{Ivezi{\'c}} {et~al.}(2008){Ivezi{\'c}}, {Sesar}, {Juri{\'c}},
  {Bond}, {Dalcanton}, {Rockosi}, {Yanny}, {Newberg}, {Beers}, {Allende
  Prieto}, {Wilhelm}, {Lee}, {Sivarani}, {Norris}, {Bailer-Jones}, {Re
  Fiorentin}, {Schlegel}, {Uomoto}, {Lupton}, {Knapp}, {Gunn}, {Covey},
  {Smith}, {Miknaitis}, {Doi}, {Tanaka}, {Fukugita}, {Kent}, {Finkbeiner},
  {Munn}, {Pier}, {Quinn}, {Hawley}, {Anderson}, {Kiuchi}, {Chen}, {Bushong},
  {Sohi}, {Haggard}, {Kimball}, {Barentine}, {Brewington}, {Harvanek},
  {Kleinman}, {Krzesinski}, {Long}, {Nitta}, {Snedden}, {Lee}, {Harris},
  {Brinkmann}, {Schneider}, \& {York}}]{Ivezic2008ApJ}
{Ivezi{\'c}}, {\v Z}., {Sesar}, B., {Juri{\'c}}, M., {et~al.} 2008, \apj, 684,
  287

\bibitem[{{Juri{\'c}} {et~al.}(2008){Juri{\'c}}, {Ivezi{\'c}}, {Brooks},
  {Lupton}, {Schlegel}, {Finkbeiner}, {Padmanabhan}, {Bond}, {Sesar},
  {Rockosi}, {Knapp}, {Gunn}, {Sumi}, {Schneider}, {Barentine}, {Brewington},
  {Brinkmann}, {Fukugita}, {Harvanek}, {Kleinman}, {Krzesinski}, {Long},
  {Neilsen}, {Nitta}, {Snedden}, \& {York}}]{Juric2008ApJ}
{Juri{\'c}}, M., {Ivezi{\'c}}, {\v Z}., {Brooks}, A., {et~al.} 2008, \apj, 673,
  864

\bibitem[{{Karlsson}(2006)}]{Karlsson2006ApJL}
{Karlsson}, T. 2006, \apjl, 641, L41

\bibitem[{{Keller} {et~al.}(2007){Keller}, {Schmidt}, {Bessell}, {Conroy},
  {Francis}, {Granlund}, {Kowald}, {Oates}, {Martin-Jones}, {Preston},
  {Tisserand}, {Vaccarella}, \& {Waterson}}]{Keller2007PASA}
{Keller}, S.~C., {Schmidt}, B.~P., {Bessell}, M.~S., {et~al.} 2007,
  Publications of the Astronomical Society of Australia, 24, 1

\bibitem[{{Kirby} {et~al.}(2008){Kirby}, {Simon}, {Geha}, {Guhathakurta}, \&
  {Frebel}}]{Kirby2008ApJ}
{Kirby}, E.~N., {Simon}, J.~D., {Geha}, M., {Guhathakurta}, P., \& {Frebel}, A.
  2008, \apjl, 685, L43

\bibitem[{{Klement} {et~al.}(2009){Klement}, {Rix}, {Flynn}, {Fuchs}, {Beers},
  {Allende Prieto}, {Bizyaev}, {Brewington}, {Lee}, {Malanushenko},
  {Malanushenko}, {Oravetz}, {Pan}, {Re Fiorentin}, {Simmons}, \&
  {Snedden}}]{Klement2009ApJ}
{Klement}, R., {Rix}, H., {Flynn}, C., {et~al.} 2009, \apj, 698, 865

\bibitem[{{Lee} {et~al.}(2008{\natexlab{a}}){Lee}, {Beers}, {Sivarani},
  {Allende Prieto}, {Koesterke}, {Wilhelm}, {Re Fiorentin}, {Bailer-Jones},
  {Norris}, {Rockosi}, {Yanny}, {Newberg}, {Covey}, {Zhang}, \&
  {Luo}}]{Lee2008AJ_SSPP1}
{Lee}, Y.~S., {Beers}, T.~C., {Sivarani}, T., {et~al.} 2008{\natexlab{a}}, \aj,
  136, 2022

\bibitem[{{Lee} {et~al.}(2008{\natexlab{b}}){Lee}, {Beers}, {Sivarani},
  {Johnson}, {An}, {Wilhelm}, {Allende Prieto}, {Koesterke}, {Re Fiorentin},
  {Bailer-Jones}, {Norris}, {Yanny}, {Rockosi}, {Newberg}, {Cudworth}, \&
  {Pan}}]{Lee2008AJ_SSPP2}
{Lee}, Y.~S., {Beers}, T.~C., {Sivarani}, T., {et~al.} 2008{\natexlab{b}}, \aj,
  136, 2050

\bibitem[{{Majewski} {et~al.}(2004){Majewski}, {Ostheimer}, {Rocha-Pinto},
  {Patterson}, {Guhathakurta}, \& {Reitzel}}]{Majewski2004ApJ}
{Majewski}, S.~R., {Ostheimer}, J.~C., {Rocha-Pinto}, H.~J., {et~al.} 2004,
  \apj, 615, 738

\bibitem[{{Mu{\~n}oz} {et~al.}(2006){Mu{\~n}oz}, {Carlin}, {Frinchaboy},
  {Nidever}, {Majewski}, \& {Patterson}}]{Munoz2006ApJ}
{Mu{\~n}oz}, R.~R., {Carlin}, J.~L., {Frinchaboy}, P.~M., {et~al.} 2006, \apjl,
  650, L51

\bibitem[{{Norris} {et~al.}(2007){Norris}, {Christlieb}, {Korn}, {Eriksson},
  {Bessell}, {Beers}, {Wisotzki}, \& {Reimers}}]{Norris2007ApJ}
{Norris}, J.~E., {Christlieb}, N., {Korn}, A.~J., {et~al.} 2007, \apj, 670, 774

\bibitem[{{Norris} {et~al.}(2010){Norris}, {Yong}, {Gilmore}, \&
  {Wyse}}]{Norris2010ApJ}
{Norris}, J.~E., {Yong}, D., {Gilmore}, G., \& {Wyse}, R.~F.~G. 2010, \apj,
  711, 350

\bibitem[{{Pagel} \& {Patchett}(1975)}]{Pagel1975MNRAS}
{Pagel}, B.~E.~J. \& {Patchett}, B.~E. 1975, \mnras, 172, 13

\bibitem[{{Placco} {et~al.}(2010){Placco}, {Kennedy}, {Rossi}, {Beers}, {Lee},
  {Christlieb}, {Sivarani}, {Reimers}, \& {Wisotzki}}]{Placco2010AJ}
{Placco}, V.~M., {Kennedy}, C.~R., {Rossi}, S., {et~al.} 2010, \aj, 139, 1051

\bibitem[{{Prantzos}(2003)}]{Prantzos2003AA}
{Prantzos}, N. 2003, \aap, 404, 211

\bibitem[{{Prantzos}(2008)}]{Prantzos2008AA}
{Prantzos}, N. 2008, \aap, 489, 525

\bibitem[{{Ryan} \& {Norris}(1991)}]{Ryan1991AJ}
{Ryan}, S.~G. \& {Norris}, J.~E. 1991, \aj, 101, 1865

\bibitem[{{Salvadori} {et~al.}(2010){Salvadori}, {Ferrara}, {Schneider},
  {Scannapieco}, \& {Kawata}}]{Salvadori2010MNRAS}
{Salvadori}, S., {Ferrara}, A., {Schneider}, R., {Scannapieco}, E., \&
  {Kawata}, D. 2010, \mnras, 401, L5

\bibitem[{{Salvadori} {et~al.}(2007){Salvadori}, {Schneider}, \&
  {Ferrara}}]{Salvadori2007MNRAS}
{Salvadori}, S., {Schneider}, R., \& {Ferrara}, A. 2007, \mnras, 381, 647

\bibitem[{{Sbordone} {et~al.}(2010){Sbordone}, {Bonifacio}, {Caffau}, {Ludwig},
  {Behara}, {Gonzalez Hernandez}, {Steffen}, {Cayrel}, {Freytag}, {Van't Veer},
  {Molaro}, {Plez}, {Sivarani}, {Spite}, {Spite}, {Beers}, {Christlieb},
  {Francois}, \& {Hill}}]{Sbordone2010astroph}
{Sbordone}, L., {Bonifacio}, P., {Caffau}, E., {et~al.} 2010, ArXiv e-prints

\bibitem[{{Schlaufman} {et~al.}(2009){Schlaufman}, {Rockosi}, {Allende Prieto},
  {Beers}, {Bizyaev}, {Brewington}, {Lee}, {Malanushenko}, {Malanushenko},
  {Oravetz}, {Pan}, {Simmons}, {Snedden}, \& {Yanny}}]{Schlaufman2009ApJ}
{Schlaufman}, K.~C., {Rockosi}, C.~M., {Allende Prieto}, C., {et~al.} 2009,
  \apj, 703, 2177

\bibitem[{{Schlegel} {et~al.}(1998){Schlegel}, {Finkbeiner}, \&
  {Davis}}]{Schlegel1998ApJ}
{Schlegel}, D.~J., {Finkbeiner}, D.~P., \& {Davis}, M. 1998, \apj, 500, 525

\bibitem[{{Sch{\"o}rck} {et~al.}(2009){Sch{\"o}rck}, {Christlieb}, {Cohen},
  {Beers}, {Shectman}, {Thompson}, {McWilliam}, {Bessell}, {Norris},
  {Mel{\'e}ndez}, {Ram{\'{\i}}rez}, {Haynes}, {Cass}, {Hartley}, {Russell},
  {Watson}, {Zickgraf}, {Behnke}, {Fechner}, {Fuhrmeister}, {Barklem},
  {Edvardsson}, {Frebel}, {Wisotzki}, \& {Reimers}}]{HESstellarV}
{Sch{\"o}rck}, T., {Christlieb}, N., {Cohen}, J.~G., {et~al.} 2009, \aap, 507,
  817

\bibitem[{{Schuster} {et~al.}(2004){Schuster}, {Beers}, {Michel}, {Nissen}, \&
  {Garc{\'{\i}}a}}]{Schuster2004AA}
{Schuster}, W.~J., {Beers}, T.~C., {Michel}, R., {Nissen}, P.~E., \&
  {Garc{\'{\i}}a}, G. 2004, \aap, 422, 527

\bibitem[{{Searle} \& {Sargent}(1972)}]{Searle1972ApJ}
{Searle}, L. \& {Sargent}, W.~L.~W. 1972, \apj, 173, 25

\bibitem[{{Spite} \& {Spite}(1982)}]{spite1982AA}
{Spite}, F. \& {Spite}, M. 1982, \aap, 115, 357

\bibitem[{{Talbot} \& {Newman}(1977)}]{Talbot1977ApJS}
{Talbot}, Jr., R.~J. \& {Newman}, M.~J. 1977, \apjs, 34, 295

\bibitem[{{Tinsley}(1980)}]{Tinsley1980}
{Tinsley}, B.~M. 1980, Fundamentals of Cosmic Physics, 5, 287

\bibitem[{{Wisotzki} {et~al.}(1996){Wisotzki}, {Koehler}, {Groote}, \&
  {Reimers}}]{Wisotzki1996AA}
{Wisotzki}, L., {Koehler}, T., {Groote}, D., \& {Reimers}, D. 1996, \aaps, 115,
  227

\bibitem[{{Yi} {et~al.}(2001){Yi}, {Demarque}, {Kim}, {Lee}, {Ree}, {Lejeune},
  \& {Barnes}}]{Yi2001ApJS}
{Yi}, S., {Demarque}, P., {Kim}, Y., {et~al.} 2001, \apjs, 136, 417

\bibitem[{{York} {et~al.}(2000){York}, {Adelman}, {Anderson}, {Anderson},
  {Annis}, {Bahcall}, {Bakken}, {Barkhouser}, {Bastian}, {Berman}, {Boroski},
  {Bracker}, {Briegel}, {Briggs}, {Brinkmann}, {Brunner}, {Burles}, {Carey},
  {Carr}, {Castander}, {Chen}, {Colestock}, {Connolly}, {Crocker}, {Csabai},
  {Czarapata}, {Davis}, {Doi}, {Dombeck}, {Eisenstein}, {Ellman}, {Elms},
  {Evans}, {Fan}, {Federwitz}, {Fiscelli}, {Friedman}, {Frieman}, {Fukugita},
  {Gillespie}, {Gunn}, {Gurbani}, {de Haas}, {Haldeman}, {Harris}, {Hayes},
  {Heckman}, {Hennessy}, {Hindsley}, {Holm}, {Holmgren}, {Huang}, {Hull},
  {Husby}, {Ichikawa}, {Ichikawa}, {Ivezi{\'c}}, {Kent}, {Kim}, {Kinney},
  {Klaene}, {Kleinman}, {Kleinman}, {Knapp}, {Korienek}, {Kron}, {Kunszt},
  {Lamb}, {Lee}, {Leger}, {Limmongkol}, {Lindenmeyer}, {Long}, {Loomis},
  {Loveday}, {Lucinio}, {Lupton}, {MacKinnon}, {Mannery}, {Mantsch}, {Margon},
  {McGehee}, {McKay}, {Meiksin}, {Merelli}, {Monet}, {Munn}, {Narayanan},
  {Nash}, {Neilsen}, {Neswold}, {Newberg}, {Nichol}, {Nicinski}, {Nonino},
  {Okada}, {Okamura}, {Ostriker}, {Owen}, {Pauls}, {Peoples}, {Peterson},
  {Petravick}, {Pier}, {Pope}, {Pordes}, {Prosapio}, {Rechenmacher}, {Quinn},
  {Richards}, {Richmond}, {Rivetta}, {Rockosi}, {Ruthmansdorfer}, {Sandford},
  {Schlegel}, {Schneider}, {Sekiguchi}, {Sergey}, {Shimasaku}, {Siegmund},
  {Smee}, {Smith}, {Snedden}, {Stone}, {Stoughton}, {Strauss}, {Stubbs},
  {SubbaRao}, {Szalay}, {Szapudi}, {Szokoly}, {Thakar}, {Tremonti}, {Tucker},
  {Uomoto}, {Vanden Berk}, {Vogeley}, {Waddell}, {Wang}, {Watanabe},
  {Weinberg}, {Yanny}, \& {Yasuda}}]{York2000AJ}
{York}, D.~G., {Adelman}, J., {Anderson}, Jr., J.~E., {et~al.} 2000, \aj, 120,
  1579

\bibitem[{{Yoshii}(1981)}]{Yoshii1981AA}
{Yoshii}, Y. 1981, \aap, 97, 280

\bibitem[{{Zhao} {et~al.}(2006){Zhao}, {Chen}, {Shi}, {Liang}, {Hou}, {Chen},
  {Zhang}, \& {Li}}]{Zhao2006ChJAA}
{Zhao}, G., {Chen}, Y., {Shi}, J., {et~al.} 2006, Chinese Journal of Astronomy
  and Astrophysics, 6, 265

\end{thebibliography}







\end{document}